# Product Differentiation and Geographical Expansion of Exports Network at Industry level*

Xuejian Wang, Zhejiang University of Finance and Economics

**Abstract**

Industries can enter one country first, and then enter its neighbors' markets. Firms in the industry can expand trade network through the export behavior of other firms in the industry. If a firm is dependent on a few foreign markets, the political risks of the markets will hurt the firm. The frequent trade disputes reflect the importance of the choice of export destinations. Although the market diversification strategy was proposed before, most firms still focus on a few markets, and the paper shows reasons.

In this paper, we assume the entry cost of firms is not all sunk cost, and show 2 ways that product heterogeneity impacts extensive margin of exports theoretically and empirically. Firstly, the increase in product heterogeneity promotes the increase in market power and profit, and more firms are able to pay the entry cost. If more firms enter the market, the information of the market will be known by other firms in the industry. Firms can adjust their behavior according to other firms, so the information changes entry cost and is not sunk cost completely. The information makes firms more likely to entry the market, and enter the surrounding markets of existing markets of other firms in the industry. When firms choose new markets, they tend to enter the markets with few competitors first.

Meanwhile, product heterogeneity will directly affect the firms' network expansion, and the reduction of product heterogeneity will increase the value of peer information. This makes firms more likely to entry the market, and firms in the industry concentrate on the markets.

Keyword: trade network, monopolistic competition, industry size, industry trade

Introduction

Trade disputes between countries have had an impact on international trade, and the choice of export destinations can impact firms' response to trade disputes. If a firm concentrate on one foreign market, the international political risks will threaten the firm. However, firms with many export markets can be less affected by international conflicts. Though the Ministry of Commerce proposed a market diversification strategy as early as the late 1990s, Chinese firms still concentrate on a few developed markets in the past 20 years. This is caused by many reasons. Among them, the successful experience of exporting to developed countries is one of the important factors. The successful experience of exporting to a certain country in history can inspire other firms in the industry.

The research on the entry decision of foreign markets can change the decisions of firms in reality. The entrance decision of firms is influenced by entry costs. While the paper can answer why firms focus on traditional developed countries, it can also help the government help firms enter new export markets. If firms can export in many different foreign markets, they can not only avoid political risks but also resist economic fluctuations (Zhang 2014, Hummels and Klenow 2005).

In international trade, social networks build up firm trade networks, and this connection helps firms explore new markets. Other study has shown that the networks structure of firm to enter foreign markets (Chaney, 2014). Network of export means that a firm is more likely to enter new markets which are close to existing markets. This is because network can reduce information costs (Bailey et al. 2020, Chaney 2014), firms can get information more easily, and this can reduce the risks (Rauch 2001), so this can benefit firms to enter new markets. The networks can reflect sequence of exports, the number of foreign markets which a firm enter. This can reflect the geography of foreign markets and meaning of distance (Chaney 2014). The network show the information of markets, the information can be delivered by importers (Erbahar 2019), as well as migrants (De Lucio et al. 2016, Gao, Huang, and Li 2019, Yang and Li 2016).

In addition to the above information sources, our model points out that geographical expansion at industry level: industry can enter one market first and then enter the its neighbors' markets. Firms in the industry can enter new markets by following other firms in the industry, and enter new markets close to existing markets of other firms. Firstly, firms can directly search for new markets by themselves, and use existing export markets to explore nearby markets. Secondly, firms to obtain information about other firms on foreign markets, and they can enter new markets of their peers, they can enter the new markets close to existing markets of their peers. The information from peers is reflected by the number of firms enter new markets, the increase in product heterogeneity can lead to an increase in the number of firms entering a market in the industry. Meanwhile, when entering the market, they face fierce competition, which makes firms willing to enter new markets where their peers are relatively less. In addition, if the firms in the industry produce similar products, the value of information is high. The similar products can encourage firms to follow their peers, and firms concentrate on certain markets.

In literature, trade network is often used to explain the meaning of distance in different models, and different models have different interpretations of the economic meaning of distance of international trade. The model of network of trade (Chaney, 2014) use different search models to explain the meaning of distance. The firm self-generation network model (Wu and Yang 2019) holds that the firm can form networks, the existing products and markets can impact enter of new markets. However, these trade network models either do not include product heterogeneity and industrial factors, or do not analyze the entry cost deeply, and their explanation of the meaning of distance is incomplete. The firm self-generation network model includes product factors (Wu and Yang 2019), but it lacks a theoretical analysis of changes in the entry cost, nor does it include the influence of industry factors. There are also studies that reveal the impact of interaction between peers on trade (Zhang, Wang, and Li 2016), but it do not analyze trade networks and entry cost. Although the model of network structure of trade (Chaney 2014) shows how firm search new markets to avoid barrier of distance, it does not analyze the factors of product and industry. The impact of product of heterogeneity on trade networks of country level is revealed before (Rauch 1999), the impact from industry level and firm level is still not revealed.

This paper shows how product heterogeneity and peer effect impact network structure of export at industry level. Learn from network structure of trade model (Chaney 2014), research on entry cost (Ruhl and Willis 2017) and MO model (Melitz and Ottaviano 2008), this paper builds up new model to research how network of exports are impacted by peers. From the aspect of impact of product heterogeneity on trade, this model expand the country level model (Rauch 1999). In addition, the entry cost assumption of the differentiation between entry cost and sunk cost is influenced by other research (Ruhl and Willis 2017). The paper assumes that the entry cost for firms can impact of peers, and assumes that the cost of entering the market is not all sunk costs. This means that after entering the market, firms can continue developing the market and deepen their relationship with peers based on changes in the market environment. Since firms will consider their future expansion behavior before entering the market, these factors will affect the firm's decision to enter the market.

In a word, the model in this paper shows 2 ways that product heterogeneity impacts industry export network and extensive margin of exports theoretically and empirically. Firstly, product heterogeneity can change the number of firms in the market, and this reflect the information of the markets and change the entry decision. The paper adds changes in the entry cost based on the MO model to show the conclusion. Secondly, the product heterogeneity can impact the entry of the markets directly. If the firms in the industry produce similar products, they are more likely to mimic other firms to build up trade networks.

In theory, firm-level trade networks are affected by product heterogeneity and industrial factors, the trade network is built up by information. The information is shown as the number of firms in an industry in the market, and this is affected by product heterogeneity. The increase in product heterogeneity makes firms have more market power. As a result, firms have the profits to pay the entry cost. The entry cost is difficult to be payed because the farther the distance is, the more difficult it is to enter the market. Once a firm enters the market, other firms in the industry will obtain this information, and they will consider whether to follow to enter the market, or explore new markets around the existing market of other firms. When a large number of firms enter this market, the profit of a typical firm is reduced, and this reflects fierce competition and firms are reluctant to enter the markets.

Meanwhile, the trade network is also affected by product heterogeneity directly. If product heterogeneity declines, the value of peer information will be higher, and firms will mimic other firms' behavior. This makes firms more willing to enter the market by following others, and firms concentrate on a few markets which are close with each other.

Empirically, the export data of Chinese firms from 2001 to 2006 robustly supports the theory above, the number of firms exporting to a certain market significantly affects the export of other firms. This influence is equivalent to the influence of firms using their existing markets to search for new markets. Meanwhile, the product heterogeneity is calculated according to the SITC three-digit code (Rauch 1999), it can positively affects the number of export firm in the industry. In addition, the empirical model uses data from 8,626 Chinese firms and 164 countries. The dependent variable is whether the firm exports or not, and the firms export probability is estimated using the Probit model. The model adds control variables, time fixed effects and industry fixed effects to eliminate the influence of other factors. Country and region fixed effects and extreme values do not affect the conclusions of the model.

The second part of the paper introduces theoretical models, then stylized facts is shown, and then empirical analysis and conclusions are shown.

1、Theory of geographical expansion of export networks

Firstly, model shows assumption of consumption, production and how peers can reduce information costs, and analyzes how number of firms entering a certain market impacts entry decision of other firms. The increase in product heterogeneity in the industry will make firms have more profits, more firms can pay entry costs and enter the markets. Secondly, this model analyzes how a firm expands use information from others to expand export network at industry level. The expansion is presented at network structure: industry can enter one market first and then enter its neighbors. Finally, given analysis above, the heterogeneity of products can also directly affect the expansion of the industry network. This is because the reduction of product heterogeneity can increase the value of information from other firms.

## 1.1 Expansion of trade network

### 1.1.1 Choice of consumer and firm

In consumption function (1.1), $p_i$ and $q_i^c$ represent price and consumption of product i individually, and $q_0^c$ is consumption of numeraire. The second term represents the utility obtained by the consumer product i. Then, when the utility brought by the M products are summed, it is obvious that the parameter α is positive. The last two items indicate that consumers prefer to consume different types of products. This means that consumers want to differentiate choices for certain products. Among them, η (positive) indicates quantity of consumer consumption is limited. Coefficient $\gamma_j$ (positive) in this model indicates the degree of product heterogeneity of goods in a certain industry. When the product heterogeneity in industry j is higher (higher $\gamma_j$), the products is unique, consumer can get more utility in the industry given quantity. Utility of products like water is quite limited per unit (although utility is larger when amount is larger), but unity of products (smart phone) is high per unit. Although the function of consumption form is similar with MO model (Melitz and Ottaviano 2008), the meaning of coefficients are different.

$$U = q_0^c + \alpha \int_0^M q_i^c di - \frac{1}{2\gamma_j}\int_0^M q_i^{c2} di - \frac{1}{2}\eta \left(\int_0^M q_i^c di\right)^2 \tag{1.1}$$

The demand function of market is equation (1.2). The number of firms in industry j is N, L represents the number of consumers, and $\bar{p}$ is the average price of products in industry j.

$$q_i = \frac{1}{\eta N + 1/\gamma_j}\left(\alpha L + \eta N\ \bar{p}L\gamma_j\right) - L\gamma_j p_i \tag{1.2}$$

If the product heterogeneity is higher, the firm can react more when price change more. The firm only inputs labor in production (1.3). In the short run, after inputting the fixed cost of entering the market, the firm make decisions according to relationship between price (p) and output (q).

$$q(c) = L\gamma_j(p(c) - c) \tag{1.3}$$

### 1.1.2 The impact of product heterogeneity on number of firms in the industry

The model added the product heterogeneity in industry ($\gamma_j$) and number of firms in the industry, and the results are the same with MO model. In the long term, the firm with the highest cost in industry j gets zero profit after paying fixed costs ($f_e$).

Meanwhile, the number of firms is affected by many other factors. In particular, the entry and exit of firms are affected by the degree of product heterogeneity.

In the long run, the network expansion behavior of firms is accompanied by the entry and exit, and the profit of firms with highest cost get zero profit. The cost of foreign firms entering the market is divided into two parts, one part is sunk cost, the other part is variable cost $I_3$. Because the sunk cost cannot be recovered, it will not affect the optimal results of the firms, which is consistent with the traditional assumption that the firms enters the market (Melitz 2003).

However, variable part of entry cost $I_3$ is not sunk cost, and firms pay 1 variable part of entry cost when they pay （$1+\delta$） unit entry cost. This means that firms can change its additional input after the firm enters the market. First, the variable entry cost $I_3$ is affected by the number of firms. The more the number of peers, the lower the cost of entering information. Changes in the number of peers will affect the cost of information. This in turn affects whether and when firms enter the market. The entry cost of peers is different from the fixed sunk cost.

Secondly, after entering the market, firms can still increase or decrease the use of information of other firms, which will also affect the cost of firms. For example, firms can increase their investment in using the information, actively participate in activities such as chambers of commerce, learn from peer experience or get more former employees from other firms. This investment can benefit the firm. This investment will affect the results of the firm's optimal choice, so the variable entry cost $I_3$ is included in the analysis of cost of foreign firms (1.4) (1.5).

$$\pi = f_e + \delta I_3 \tag{1.4}$$

$$I_3 = \sigma(ie^{-N_s+1} d_s) + (1-\sigma)\left(d_{0s} + d_s^{'} - d_s^p\right) \tag{1.5}$$

Variable part of entry cost $I_3$ contains the cost of entering market by following other firms and entry new market around existing markets of other firms. The model assumes that the relative ratio of firms which enter the markets by following others is $\sigma$。 The two search models above are discussed in next section. Finally, the variable cost of firms is still affected by factors such as price and products (1.7). This the product of equation (1.7) and $s_s$. The ratio $s_s$ is smaller than 1, this means than foreign firms have lower costs to enter the market, because they have to pay the variable part of the entry cost. The number of firms exports to country s within an industry（$N_s$）is part of the total number of firms within the industry（N）, the ratio of $N_s$ to N is $x_s$. The sum of entry cost and other variable cost is cost of zero profit firms (1.6). The profit of firm is equation (1.8).

$$c_{Ds} = I_3 + I_4 \tag{1.6}$$

$$I_4 = \frac{s_s}{\eta N_s + 1/\gamma_j}\left(\frac{\alpha}{\gamma_j} + \eta N_s \, \bar{p}\right) \tag{1.7}$$

$$\pi(c) = \frac{L\gamma_j}{4}(c_D - c)^2 \tag{1.8}$$

Therefore, when the market is in a long-term equilibrium, the equilibrium number of firms in a certain industry in a market will increase when product heterogeneity increase (1.9). To get this, the model assume that distance between exporters and importers (existing markets) $d_{0s}$ (or $d_s$) is big enough, and N is big integer. The proof can be found in appendix.

$$\lim_{d_{0s} \to \infty} \frac{\partial N_s}{\partial \gamma_j} = \frac{\frac{1}{\gamma_j^2}\sigma i e^{-N_s+1} + \frac{(1-\sigma)}{\gamma_j^2}}{-\sigma i e^{-N_s+1}\left(\eta N_s - \eta + \frac{1}{\gamma_j}\right) + (1-\sigma)\gamma_j\eta} \approx \frac{1}{\gamma_j^3 \eta} > 0 \tag{1.9}$$

The assumption that $d_{0s}$ is really big means that firms search efficiently by itself and the obstacle of distance is relatively big. Therefore, firms may search far enough markets to act as an intermediary for its own expansion, this can be supported in previous research (Chaney 2014). In addition, N is obviously an integer, and N could be a relatively big integer because it stands for the number of firms in one industry. The data in empirical model shows that the median and mean of N is 5 and 28 individually. In that case, it is big enough to make $N_s e^{-N_s+1}$ equals 0.09 and $5.3 \times e^{-11}$, and this make denominator positive (1.9).

Equation (1.9) means that the number firms increase when product heterogeneity increases. This is because similar products within the industry make the scale of peers small. When the distance to search for foreign markets is big enough, the high cost of entry prevents firms from entering the international market. Therefore, the cost of entering the market is relatively high, the number of exporting firms is small.

**Proposition 1** When product heterogeneity in the industry is high, the competition among firms is low and the market power is higher. Firms with higher market power can pay the high cost of entering foreign markets, which increases the number of firms entering this market.

## 1.2 Expansion of trade network in the industry

### 1.2.1 Decision of exports

After maximizing profits in the domestic market, firm will think about entering overseas markets. However, for firms, it is hard to enter overseas markets. Firms lack sufficient information, need to pay additional sales costs, and may also face difficulty from culture and institution. However, considering randomness faced by the firms, the firm may suddenly acquire overseas information or get new technology, and they are able to pay the cost of entering foreign markets.

In addition, the previous entry costs of foreign markets are sunk costs, which can be regarded as part of the fixed costs. In this way, the optimization decision-making process of the previous export-oriented enterprises is similar to that of domestic firms. To simplify the analysis, the model does not consider the impact of tariffs. The firm want to enter foreign market for new markets ($\Delta L_s$). At time t, the firm expands to the new market (s), as long as it meets one of the following two conditions (1.10) (1.11).

$$\Delta \pi_{ts}^l = \frac{\Delta L_s \gamma_j}{4} \left(c_D - c\right)^2 - id_s + \mu_t > \pi_{s,t-1} \qquad (1.10)$$

$$\Delta \pi_{ts}^r = \frac{\Delta L_s \gamma_j}{4} \left(c_D - c\right)^2 - id_s' - d_{0s} + v_t > \pi_{s,t-1} \qquad (1.11)$$

$$d_s < d_s' + d_{0s}, \; i > 1 \qquad (1.12)$$

The two conditions are conditions for local search and remote search. In this paper, we analyze the search model impacted by production and consumption, the profit for local search and remote search are equation (1.10) and (1.11). In equation (1.10), $d_s$ is the distance from importers to exporters, and ids stands for the information cost of exporters to get markets. The unit cost of information is i (i>1), $\mu$ is random error. In profit from remote search (1.11), $d_{0s}$ stands for distance between exporters to existing partners of exports, $d_s'$ stands for distance between existing partners to new markets.

So, when does a firm conduct a local search rather a remote search? This depends on the unit information cost i and the distance between the exporter and importer. In terms of distance, the straight-line distance from the exporters to importers is less than the broken line distance through the third country (1.12). However, this distance effect will be important if the gap is large.

Generally speaking, when the export market of the firm is far away, the cost of opening up the market brought by the remote search will be less than the cost of the local search. Although the straight-line distance of the local search is shorter, the unit cost of the remote search is lower. This is because the remote search is between existing partners and new markets, and information cost is low (i>1). In addition, firms can generally find a route with lower cost than local search, because it may have many potential remote search routes.

**Proposition 2** The information cost of entering the market makes firms to enter the near market firstly. When the market is far away, firms can use remote search to reduce the unit distance information cost of entering a new market. The cost of entering the market by remote search may be lower than the cost of local search. As a result, firms tend to search partners in neighboring countries directly, use existing partners to find new partners far away, and expand the trade network.

### 1.2.2 Following others

For the first time, this model defines the following others of trade network expansion in the industry, and this means after other firms successfully export to a certain country, the firms also export to that country. The market following is the first step to expand the trade networks. This means that the information of a certain firm's success in a certain overseas market spreads in the home country. This information can be observed by others, or the firms can obtain this information through the flow of labor in the labor market. This can reduce the cost of firms directly to enter the market. After the cost is reduced, given production cost of zero profit firms, the profit of firms which enter the markets by following others is equation (1.13).

$$\Delta \pi_{ts}^{f} = \frac{\Delta L_S \, \gamma_j}{4} (c_{Ds} - c)^2 - \delta i e^{-N_s + 1} \, d_s + \mu_t \qquad (1.13)$$

From the perspective of the entry cost, the number of firms whose industries export to a certain country ($N_s$) increases, the cost of firm enter the market by following others decreases. The decrease function is not a linear function, and the decrease is quite sharp at first. When the number of firms becomes 1, this search mode degenerates into direct search by firms.

Model assume that if a firm pay (1+δ) to entry the market, the sunk cost is only δ. The sunk cost cannot interfere the decision of firm. However, part of the entry cost is not sunk cost, this means the part of entry cost can be refundable. This part of the additional investment represents the cost of firms constantly looking for new partners and sales channels, and is a reflection of the variable costs of expanding the market after the firm enters the market. This part of the cost is discussed in section 1.2.

1.2.3 Search for new markets using others

Using the export information of other firms in the country, the firm can not only consider exporting to the country, but also export to its neighboring countries. Because others enter the neighboring country of country s, the firm enters the market of country s, and distance between country s and its neighbor is $d'_{is}$. The difficulty in entering country s can be shown by the sum of distance between country s and every existing market in the industry. The profit is shown in equation（1.14）（1.15）（1.16）.

$$\Delta \pi_t^n = I_1 - I_2 \tag{1.14}$$

$$I_1 = \frac{\gamma_j \Delta L_s}{4} (c_{Ds} - c)^2 \tag{1.15}$$

$$I_2 = \delta(d_{0\,s} + d'_s - d_s^\rho) + v_t, \quad d'_s = \Sigma_i d'_{is}, \Sigma_i e^{-N_i/L_i + 1} d'_{is} = d_s^\rho \tag{1.16}$$

In addition, there is an interactive effect between the number of firms in the markets and the distance from the market and to every market in the industry. The model divides the distance by the density of firms in the industry market to measure this interaction effect (1.16). The market has a high density of firms in a certain industry, and this means that the number of firms entering the market per unit of GDP is large, and the distance between different markets is still very close. These means that there is fierce competition in these areas. The increase in distance and the decrease in density reduce competition and reduce the barrier to trade.

If a firm choose to enter the market by following others or enter nearby markets of existing markets of others in the industry, the cost of a firm to enter new markets will decrease as the number of export firms in the industry increases. The equation (1.17) shows the effects the number of firms on following, and the impact of using others to search for new markets is similar. The number of peers in the market can reduce the information cost of others. They use this information to follow or expands to the neighboring countries of this country, and the expansion activities of all firms in the industry form an industrial expansion. If there is only one firm in the industry, then new market search in this industry degenerates into the remote search of the firm.

$$\frac{\partial \Delta \pi_t^f}{\partial N_s} = ie^{-N_s + 1} d_s \delta > 0 \tag{1.17}$$

Similarly, when a firm explores new markets, if the existing market is close to the new market, the information cost will be lower (1.18). This reflects the closeness of between the adjacent market and the original market. If markets are more similar, information about similar markets will be more valuable. Meanwhile, if distance is small, it will be easier to obtain information about new markets. In addition, when the density of firms in certain markets decreases, this makes enterprises more willing to expand given distance among markets. This indicates impact from competition.

$$\frac{\partial \Delta \pi_t^n}{\partial I_2} \frac{\partial I_2}{\partial d_s'} = -\delta < 0 \tag{1.18}$$

$$\frac{\partial \Delta \pi_t^n}{\partial I_2} \frac{\partial I_2}{\partial d_s^\rho} = \delta > 0 \tag{1.19}$$

**Proposition 3** If the number of export firms in the industry increases, the industry's information on foreign markets will also increase. This promotes the reduction of entry costs. Therefore, firms can pay for the entry cost by following, or enter nearby markets of existing markets of others.

**Proposition 4** Given certain market distances, when there are relatively less firms in certain markets, firms face less competition and are willing to enter the markets.

## 1.3 Impact of variable cost and value of information from peers

The number of industries N will not only provide the export information of firms, but also change highest cost of firms surviving in the market ($c_D$). The costs will affect the firms' entry decision. In particular, the costs will change due to product heterogeneity. First, the analysis above needs to be based on the impact of firms' output and cost.

### 1.3.1 Impact from firm's output and cost

To start with, the expansion of the export network is affected by the output of the firm. The profit earned by the firm export is obviously positively related to its own production capacity (1.20).

$$\Delta\pi_{ts}^n = \frac{\Delta q_s}{2}(c_{Ds} - c) - \delta\left(d_{0s} + d_s' - d_s^\rho\right) - v_t \qquad (1.20)$$

The impact of output on firm's profits is influenced by highest costs of firms surviving in the markets ($c_D$). When $c_D$ is higher, the profit per unit output of the firm is higher, and the profit and the probability of entering the market is higher (1.27). This is because when the $c_D$ is higher, the cost gap between $c_D$ and the cost of the firm is bigger. Next, the model focuses on analyzing the impact of other factors on the zero-profit cost ($c_D$), and these factors influence firms' entry decision.

### 1.3.2 The impact of value of information from peers

When firms produce similar products, the information from other firms is very valuable. This is because when the products are similar, other firms can copy the successful experience of others. They face similar problems, and they produce and sell similarly. The impact of value of information on following other to export is equation (1.21), and the impact of searching new markets around existing markets of others is equation (1.22).

$$\frac{\partial^2 c_{Ds}}{\partial \gamma_j \partial N_s} = \frac{-s_s}{(\eta N_s \gamma_j + 1)^4}((\alpha\eta - \eta\overline{p}))(-\eta^2 N_s^2 \gamma_j^2 + 1) < 0, \qquad (1.21)$$

$$\frac{\partial^2 c_{Ds}}{\partial \gamma_j \partial d_s^\rho} = \frac{\partial^2 c_{Ds}}{\partial \gamma_j \partial N_s}\frac{dN_s}{dd_s^\rho} = \frac{1}{\Sigma_i e^{-N_i/L_i+1}/L_i d_{is}'} \frac{s_s}{(\eta N_s + \gamma_j)^4}((\alpha\eta - \eta\overline{p}))(-\eta^2 N_s^2 \gamma_j^2 + 1) > 0 \qquad (1.22)$$

When the value of information of others in the industry is higher (lower $\gamma_j$), the zero-profit cost to survive in the markets ($c_D$) increases, and other firms want to enter the markets. This means that when reference value between firms is high, the value by copying other firms' behavior is high, and the firms can obtain more revenue. However, the free entry of the market cuts the profit of firms with cost $c_D$ to zero. Firms with high costs in the new market can also survive. For most firms, the profit per unit product of firms rises, and firms are more willing to export. The higher profit makes the firms enter the markets by following others. In addition, when the value of information among firms is higher (lower $\gamma_j$), firms focus on a few markets which are close with each other (lower $d_s^\rho$) (1.22).

**Proposition 5** Low product heterogeneity in the industry makes value of information among firms high, firms are more willing to follow the market, and firms concentrate on a few markets which are close with each other.

In short, firms can entry new markets by following the market, and enter new markets around existing markets of other firms. First, firms can directly search for new markets by themselves, and firms can also use their own existing export markets to explore new markets around existing markets (Proposition 2), and Chaney (2014) has shown this one. In this model, the product characteristics of different industries can affect the industry's overseas expansion. When product heterogeneity increases, firms which enter a certain market in the industry increase (Proposition 1). This can increase the information about foreign markets in the industry, which in turn encourages firms in the industry to enter the markets and new markets around the existing markets of other firms (Proposition 3). When firms want to entry the surround markets of existing markets of others, they prefer markets with less other firms for competition (Proposition 4). In addition, the more similar the products produced by firms in the industry, the higher the value of information among firms in the industry. The high-value market information can promote firms enter markets by following others, and they focus on a few markets which are close with each other (Proposition 5).

## 2. Stylized facts

From the above theoretical model, firms can enter the market by following peers and use peers to search for new markets around the existing markets of peers. To test the theory, the data of the China Customs database from 2001 to 2006 was merged with the firm data of annual surveys by China's National Bureau of Statistics (NBS). The firm level data is discussed and sorted out in other researches (Brandt, Van Biesebroeck, and Zhang 2012, Brandt et al. 2017), the way of merging 2 data sets is based on another research (Fan, Lai, and Li 2015).

The sample size of the data is 7073320, and this is the product of 8,626 firms, 164 exporting countries and 5 years (2002-2006). All firms exported every year during the 6 years (2001-2006), but the data in 2001 was used to reflect the impact of lag. The limited firm sample can eliminate the factors that caused the firm to stop exporting completely, thus allowing the model focus on the reasons behind the network expansion of exports. From the exports data of 8626 firms from 2001 to 2006, the stylized facts can be shown below.

As shown in Figure 1, most firms export to a small number of countries, and only a very small number of firms export more than 60. The number of export destinations for most firms is concentrated within 20. This shows that there are many factors preventing firms from exporting to many destinations.

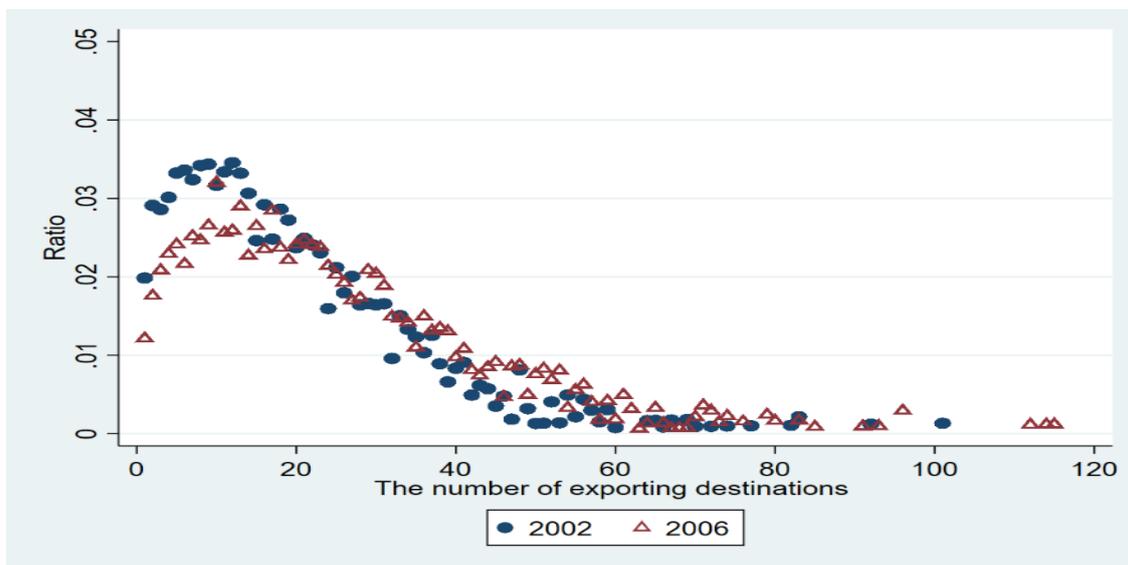

Figure 1 The number of markets which a firm enters and its frequency

Most firms have very few export markets (Table 1), and the gap in the number of export markets in different industries is also very large. There are no zero values in firms exports data, because the firms export at least once in the data set every year. However, there is a large number of zero values in the export volume of the industry to a certain country, and this shows the unbalanced exports between industries. The gap between the 95th quantile and the median is much larger than the gap between the median and the quintile.

Table1 Number of markets which a firm enters and number of firms which enter a market in an industry

| Year/Percentile | Number of markets which a firm enters | Number of firms which enter a market in an industry |
| --- | --- | --- |
| 2002/ quintile | 1 | 0 |
| 2002/ median | 4 | 0 |
| 2002/ 95th quantile | 26 | 64 |
| 2006/ quintile | 1 | 0 |
| 2006/ median | 7 | 1 |
| 2006/ 95th quantilet | 37 | 95.5 |

In addition, there is an inverted U-shaped relationship between the number of firms in an industry and the average distance from importers to China (figure 2). When the number of firms is less than 100, the average distance is positively correlated with the number of firms in an industry. This could reflect that the increase in the export information of firms has made the distance of exporters in this industry longer, and the industry shows a trend of global expansion. However, when the number of firms exceeds 300, distance is negatively correlated with number of firms. Excessive number of firms in the industry could intensify competition, and this could decrease the distance of exporters.

Figure 2 Number of firms in an industry and average distance from importers to China

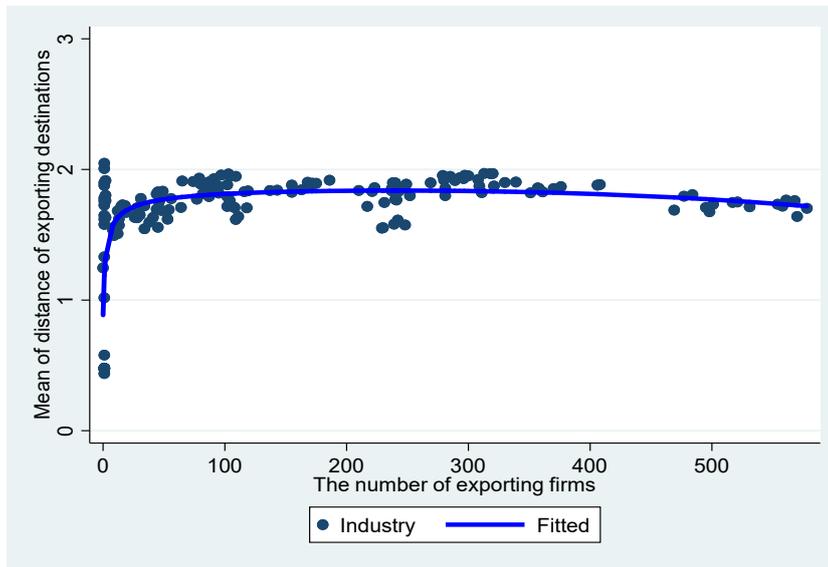

The number of markets which a firm exports to is negatively correlated with the number of firms in the industry at first, and then is positively correlated with the number of firms. This shows that there are some opposite factors that make the two variables change. The left side of the inverted U-shape indicates that as the number of firms increases, the number of markets which a firm exports to also increases. This may reflect that externalities promote the increase of markets of firms. The right side of the inverted U-shape indicates that as the number of firms increases, the number of markets that a firm export to decreases. This means that a factor hinders the large export of certain industries. This It may reflect fierce market competition.

Figure 3 Number of markets which a firm exports to and number of firms in the industry

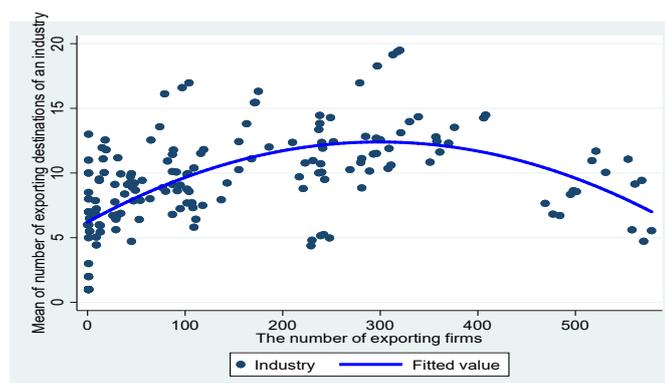

## 3. Empirical analysis of expansion of exports network in the industry

### 3.1 Empirical model

Based on empirical model in Chaney (2014), the model in this paper firstly brings up the idea of expansion of networks because of other firms in the industry. The results of this model can reflect the meaning of distance and GDP, and this can be shown by compared with the standard gravity model. The probability of firm j in industry k which export to country s is shown below, and i stands for country i and is used for sum of existing markets.

$$\Pr(export)_{sj} = \beta_0 + \beta_1 markets_j + \beta_2 \Sigma_{i,firm} d'_{isj} + \beta_3 N_{sk} + \beta_4 \Sigma_{i,industry} d'_{isk}$$
$$+ \beta_5 \Sigma_{i,industry} e^{-N_i/L_i+1} d'_{isk} + \beta_6 \gamma_k N_{sk} + \beta_7 \gamma_k \Sigma_{i,industry} e^{-N_i/L_i+1} d'_{isk} + \beta_8 \gamma_k$$
$$+ \beta_9 scale_j + \beta_{10} GDP_s + \beta_{11}\, d\ to\ CN_s + \beta_{12} \pr(export)_{sj,lag} + \beta_{13} d\ to\ world_s$$
$$+ \beta_{14} imports\ from\ CN_s + \beta_{15} imports\ from\ world_s + FE + \mu \quad (24)$$

The independent variable ($\Pr(export)_{sj}$) in the model is the export probability of firm to country s. The export status is a dummy variable. If a firm exports, the value of this variable is 1, otherwise it is 0. The model analyzes the factors which impact the dummy variable, so this model uses probit regression.

In terms of dependent variable for measuring the networks of firms, the number of markets ($markets_j$) refer to the number of markets which a firm exports to. The definition is firstly brought up by Chaney (2014). The variable is lagged for one year to reflect the lagged impact. If a firm exports to many market, then it could use them to gather more information. This can become the foundation of exports network, so its coefficient is expected to be positive.

In addition, the distance to other markets of firm ($\Sigma_{i,firm} d'_{is}$) is the sum of logarithm of distance from one country to any other existing markets of the firm. This is also firstly brought up by Chaney (2014). The variable is lagged for one year to show the lagged impact. When a firm's potential export market is close to the existing export market of the firm, it is easy for the firm to use the existing market to get more information, and the sign is expected to be negative.

The core explanatory variable is the variable of the industrial export network, and this paper firstly bring up three variables to measure this. First, the follow in the industry ($N_s$) is defined by the number of firms which exports to the country in the industry a year ago. The $N_s$ is not used in logarithm form because it is count data, and there are many zero observations (O'Hara and Kotze 2010). If more firms export to the country in the industry, and this can give enough export information to other firms in this industry, so other firms can enter the markets more easily. The lagged term is used for lagged impact. The impact of information from other firms could have positive impact, so the coefficient is expected to be positive.

In addition, the second term to measure industry network is the distance to other markets of industry, and this defined by the sum of logarithm of distance between the country and any other existing markets in the industry ($\Sigma_{i,industry} d'_{is}$). The distance to other markets of industry is lagged for one year, and this can reflect lagged impact. If a country is very close to other markets of a certain industry, the market information is easy to be found by firms. This information spreads among other firms in the industry, and this can promote exports. Therefore, sign of coefficients is predicted to be negative.

Finally, the third term to measure industry network is the intersection term between scarcity of firms and distance to other markets of industry ($d_s^\rho = \Sigma_{i,industry} e^{-N_i/L_i+1} d'_{is}$). This is defined by the distance to other markets of industry divided by negative exponential function of number of firms in the market per GDP, and the term is summed in any other existing markets firms in the industry. Because if the number of firms per GDP in a market for a certain industry is very low, it means that the market is less competitive in the industry, and the firms in this market are relatively sparse. The less competition could attract more firms in the near future. Therefore, the sign of coefficients is predicted to be positive.

In addition, the data on product heterogeneity of industry ($\gamma_k$) comes from previous research on product-to-trade networks (Rauch 1999). This row data is product-level, and the products are divided into homogeneous products and heterogeneous products according to the SITC three-digit code. In order to obtain the information of the homogeneous and heterogeneous products of the customs data, the HS code of trade data is converted to SITC to get product heterogeneity. The product heterogeneity products of industry is the ratio between total trade value of heterogeneous products and the total trade value.

In order to measure the interactive effect of the degree of product heterogeneity ($\gamma_k N_s$, $\gamma_k \Sigma_{i,industry} e^{-N_i/L_i+1} d'_{is}$), this model constructs interaction terms between product heterogeneity and the number of firms which exports to the country in the industry. The model also includes intersection term among product heterogeneity, scarcity of firms and distance to other markets of industry. If firms in the industry generally export homogeneous products, the information between manufacturers has a huge impact. This can encourage peers to enter the markets by following others. Coefficient of interaction terms between product heterogeneity and the number of firms is expected to be negative.

Furthermore, low product heterogeneity will make firms enter the nearby markets where many firms entered before. The sign of intersection term among product heterogeneity, scarcity of firms and distance to other markets of industry is expected to be positive.

In terms of control variables, the model controls the scale of the firm ($scale_j$) in a certain market, which is defined as the export volume of a firm in a certain market. It is expected that the larger the size of the firm, the stronger the export capacity, and the higher the probability of its expansion in the new market. Meanwhile, the export decision of the firm to the market one year ago ($\text{pr}(export)_{s,lag}$) is controlled, which can reflect the continuity of the export status of the firm. The coefficient of the export decision one year before should be positive. GDP of the market is controlled, which is the basic setting of the gravity model. Because high-income countries can buy more products, the coefficient of GDP is expected to be positive. In addition, distance from China to market s ($d\ to\ CN_s$) is controlled, and the following terms of distance in the paper are all in logarithmic form. Because the farther away a country is, the smaller its export volume. The coefficient of distance is expected to be negative. For the same reason, the coefficient of the sum of logarithm of distance between market s and any other country ($d\ to\ world_s$) is expected negative. The growth rate of logarithmic import volume from China to country s ($imports\ from\ CN_s$) is controlled, and import growth rate from any other countries to country s ($imports\ from\ world_s$) is controlled. This controls the impact of the macro impact of imports. If a country prefers imports, then this country may import more products from Chinese firms. The sign of the increase rate of imports from any other countries is expected to be positive. If a country prefers to imports from China, then the probability of it importing from a Chinese firm also rises. The coefficient of the growth rate of total import volume from China is also expected to be positive. Finally, μ is random Errors. The probability of firm can export to country s is:

In order to control more potential influencing factors, the model controls time fixed effects and industry fixed effects. Time fixed effects control variables that remain within a given time, such as technological changes and macroeconomic trends in a certain year. Correspondingly, industry fixed variables control variables that remain constant in a given industry. In this way, the model includes the technology level and labor characteristics of the industry. At the same time, the model uses the cluster standard errors at the firm level.

3.2 Data and descriptive statistics

The export data comes from the Chinese customs database. Data of firms and industries come from annual surveys by China's National Bureau of Statistics (NBS). The GDP of the importing country comes from the Penn World Table, and the distance is from the CEPII database. The data on product heterogeneity comes from previous research on product heterogeneity (Rauch 1999).

Descriptive statistics (Table 2) show that most variables have 7073320 samples, which are the product of 8626 firms, 164 countries and 5 years (2002-2006). The data in 2001 is used for lagged variables. When a firm does not export to a certain country, the export is zero instead of missing value. The 8626 firms exported at least once a year from 2001 to 2006. These continuous export firms make the model focus on the analysis of the transfer between the export destinations, rather than the factors that cause the firms choose to export or not not export at all.

Table 2 Descriptive statistics

|  | count | mean | sd | min | max |
|---|---|---|---|---|---|
| $\Pr(export)_{sj}$ | 7073320 | .0639756 | .2447095 | 0 | 1 |
| $\pr(export)_{sj,lag}$ | 7073320 | .0590013 | .2356271 | 0 | 1 |
| $markets_j$ | 7073320 | 9.676211 | 10.87743 | 1 | 119 |
| $\Sigma_{i,firm} d'_{is}$ | 7073320 | 18.02332 | 20.18749 | -6.18622 | 306.6471 |
| $N_{sk}$ | 7073320 | 27.89191 | 55.85964 | 0 | 578 |
| $\Sigma_{i,industry} d'_{is}$ | 7073320 | 215.264 | 47.90267 | -4.03476 | 398.1325 |
| $\Sigma_{i,industry} e^{-N_i/L_i+1} d'_{is}$ | 7073320 | 221.6113 | 51.62883 | -10.13144 | 582.7308 |
| $\gamma_k N_{sk}$ | 7073320 | 20.62869 | 44.31315 | 0 | 573.5986 |
| $\gamma \Sigma_{i,industry} e^{-N_i/L_i+1} d'_{is}$ | 7073320 | 155.4313 | 70.694 | -9.53648 | 569.6089 |
| $\gamma_k$ | 7073320 | .7084905 | .2721627 | 0 | 1 |
| $scale_j$ | 7073320 | 16.40453 | 2.490183 | 2.70805 | 26.58075 |
| $GDP_s$ | 7073320 | .3522438 | 1.249277 | .000289 | 15.26991 |
| $d\ to\ CN_s$ | 7073320 | 2.120571 | .5201752 | .1554342 | 2.950219 |
| $imports\ from\ world_s$ | 7073320 | .3165841 | .3737353 | -2.093058 | 3.638906 |
| $imports\ from\ CN_s$ | 7005002 | .1212942 | .3920436 | -14.64169 | 13.44151 |
| $d\ to\ world_s$ | 7073320 | 288.877 | 45.41201 | 225.4721 | 427.475 |

3.3 Baseline regression

Baseline regression shows that the industrial trade network has a significant impact on exports of firms (Table 3). In the model, all coefficients except $\gamma$ in second column are significant at a significance level of 1%. In terms of control variables, except for the distance to China and the growth rate of imports from the world, the signs of the coefficients of other control variables are in line with expectations. After adding the variables of the industrial trade network, this will change the meaning of distance to China and growth rate of total imports worldwide.

The results show that if the number of firms exporting to market s in industry increases, the probability that a firm in the industry exports to this market will increase, and firms prefer to expand around the existing market of other firms (Proposition 3). Meanwhile, given the distance among different markets, if the number of firms in a market increases, the export probability will decrease for more competition (Proposition 4). The expansion of industry is similar with expansion of firms, the firms can directly and search for new markets using its existing markets (Proposition 2). Results of Proposition 2 confirms the impact of firm trade network from French data (Chaney 2014).

Table 3 baseline regression

| | $\Pr(export)_{sj}$ | | | |
|---|---|---|---|---|
| $N_{sk}$ | 0.00881*** | 0.0111*** | 0.00702*** | |
| | (0.0000546) | (0.000163) | (0.000121) | |
| $\Sigma_{i,industry} d'_{is}$ | -0.000868*** | -0.000783*** | -0.0132*** | |
| | (0.000149) | (0.000153) | (0.000192) | |
| $\Sigma_{i,industry} e^{-N_i/L_i+1} d'_{is}$ | 0.00520*** | 0.00374*** | 0.00291*** | |
| | (0.000110) | (0.000164) | (0.000142) | |
| $\gamma_k N_{sk}$ | | -0.00280*** | -0.00195*** | |
| | | (0.000205) | (0.000143) | |
| $\gamma \Sigma_{i,industry} e^{-N_i/L_i+1} d'_{is}$ | | 0.00256*** | 0.00260*** | |
| | | (0.000200) | (0.000163) | |
| $markets_i$ | 0.113*** | 0.101*** | 0.0367*** | 0.0462*** |
| | (0.00172) | (0.00170) | (0.000927) | (0.000968) |
| $\Sigma_{i,firm} d'_{is}$ | -0.0411*** | -0.0427*** | -0.0153*** | -0.0224*** |
| | (0.000742) | (0.000762) | (0.000398) | (0.000423) |
| $scale_i$ | 0.143*** | 0.152*** | 0.142*** | 0.162*** |
| | (0.00250) | (0.00275) | (0.00252) | (0.00165) |
| $\gamma_i$ | 0.127** | -0.493*** | | |
| | (0.0621) | (0.0517) | | |
| $GDP_s$ | | 0.0313*** | 0.134*** | 0.137*** |
| | | (0.000924) | (0.000798) | (0.000846) |
| $d\ to\ CN_s$ | | 0.0582*** | -0.217*** | -0.288*** |
| | | (0.00351) | (0.00246) | (0.00255) |
| $\text{pr}(export)_{s,lag}$ | | 1.976*** | 2.197*** | 2.356*** |
| | | (0.00652) | (0.00622) | (0.00632) |
| $d\ to\ world_s$ | | 0.00818*** | 0.00317*** | 0.000970*** |

|  |  |  |  |  |  |
|---|---|---|---|---|---|
|  |  | (0.000130) | (0.0000425) | (0.0000339) |  |
| $imports\ from\ world_s$ |  | 0.0176*** | -0.0288*** | -0.0312*** |  |
|  |  | (0.00247) | (0.00194) | (0.00188) |  |
| $imports\ from\ CN_s$ |  | 0.0877*** | 0.0390*** | 0.0400*** |  |
|  |  | (0.00323) | (0.00233) | (0.00226) |  |
| N | 7073320 | 7073320 | 7004218 | 7004218 | 7004218 |
| Industry FE | Yes | Yes | Yes | Yes | Yes |
| Year FE | Yes | Yes | Yes | Yes | Yes |

\* p < 0.1, \*\* p < 0.05, \*\*\* p < 0.01. Clustered standard error controlled at firm level is in the parentheses.

The impact of the firm's own connection has significant impact on trade economically, and industrial connection firstly brought up in this paper has similar impact. Although the coefficient of $N_s$ (the number of peers in the markets) is 1/5 (the third column) of the coefficients of the number of export countries ($markets$), the standard deviation of $N_s$ is 5 times that of $markets$.

The interaction term $\gamma_k N_{sk}$ between product heterogeneity and industry firms is negative, which indicates that high product differentiation prevents firms from entering the market by following others (Proposition 5). The interaction term between product heterogeneity and the sparseness of peers in the market ($\gamma \Sigma_{i,industry} e^{-N_i/L_i+1} d'_{is}$) is positive. Given the number of firms in markets, this indicates that high products heterogeneity significantly hinder firms from entering neighboring market of the existing market of industry.

The results of the first three columns are not much different. The first column shows the results of only controlling industrial trade networks, firm trade networks, time fixed effects and industrial fixed effects. From column 1 to column 2, the absolute value of the coefficients of variables above does not change much. The second column adds product heterogeneity and its interaction term. The third column shows the results after adding control variables. Although the absolute value of the coefficient has changed because of control variables, the sign of the coefficients of each proposition remains unchanged, and the significant level does not change.

3.4 Comparison with standard gravity model

The last column shows the results of the standard gravity model (table 3), and the fourth column shows the results of removing the industry network and controlling the firm trade network. They serve as a control group, and coefficients of three variables change a lot from column 3 to column 4 or 5. This indicate that industrial trade network shows its economic impact from distance between the firm and the market, GDP and growth rate imports from the world.

Among them, by comparing the coefficient of distance to China, the absolute value in the standard gravity model coefficient is 5 times larger than that in the third column, and the sign between them is opposite.

This shows that the distance to China in the standard gravity model is greatly affected by the expansion of industrial networks, especially by other measurements of distance ($\Sigma_{i,firm}d'_{is}$ $\Sigma_{i,industry}d'_{is}$ and $\Sigma_{i,industry}e^{-N_i/L_i+1}d'_{is}$). The compassion from column 4 to 5 show the 25% smaller coefficients of distance to China, and the impact of trade network can show the impact of distance to China.

Similarly, the rest of change in coefficient of distance to China show the impact of industry network (or distance). The imitation and expansion of peers greatly explains how firms go beyond the barrier of distance to obtain trade information, and then imitate and find new partners through peers. This causes the coefficients of the gravity model's distance to China to become a positive number after adding the terms of the industrial network.

The positive coefficients of distance to China after controlling the terms of industrial trade network can be explained by institution, technology or randomness. The quality of institution will be generally better if the country is far away from China (developed countries). The developed countries with high quality of institution could trade more with China. Similarly, positive coefficients of distance to China can be explained by technology or randomness. A firm gets high technology randomly and has stronger ability to exports. The firm with technological improvements wants to trade with the major partners, and the major partners are developed countries and far away. The long distance could be a barrier for normal firms, but the firm with technological improvements can pay the cost to trade with long-distance partners.

For coefficients of GDP, absolute value of the coefficients in the standard gravity model GDP coefficient is 6 times larger than it in model with industry networks. This reflects GDP can be represented by demand of the economy, and this can be reflected by the number of firms to the market.

The sign of the coefficient growth rate of total imports from the world has changed between standard gravity model and model with industry network. This means that the industry networks could impact the imports from other countries.

3.5 Discussion of endogeneity

The core explanatory variables of the model are industry connections ($N_s$) and distance ($\Sigma_{i,industry}d'_{is}$, $\Sigma_{i,industry}e^{-N_i/L_i+1}d'_{is}$). Other important explanatory variables include product heterogeneity ($\gamma$). The explanation of the causality of the variables is reliable.

To start with, the above-mentioned important explanatory variables and the explained variables will not have reverse causality. First, the above-mentioned causal relationship is strongly supported by the model based on theory of heterogeneous trade (Melitz 2003, Melitz and Ottaviano 2008). Model settings for firms and consumers are relatively mature. Meanwhile, industry-level variables are unlikely to have reverse causal effects on individual export behaviors of firms, as long as export behaviors happen in a monopolistic competitive market. Even if individual firms have the ability to affect the entire industry, results are robust in the regression where largest firms are removed.

In addition, omitted variables will not affect the results, because many control variables and fixed effects make the results unbiased. In short, the propositions of empirical research and theoretical analysis are consistent.

3.6 Empirical analysis for mechanism

The impact of industry networks on the export of firms has been analyzed, but the mechanism needs to be empirically tested in this section. The mechanism regression of this model focuses on the impact of product heterogeneity on the number of firms in the industry (Proposition 1). The mechanism 1 can strongly support the argument that the increase in the number of firms leads to the increase in the probability of entry (Proposition 3).

$$\begin{aligned} N_{sk,now} = & \beta_0 + \beta_8 \gamma_k + \beta_1 markets_j + \beta_2 \Sigma_{i,firm} d'_{isj} + \beta_4 N_{sk} + \beta_4 \Sigma_{i,industry} d'_{isk} \\ & + \beta_5 \Sigma_{i,industry} e^{-N_i/L_i+1} d'_{isk} + \beta_9 scale_j + \beta_{10} GDP_s + \beta_{11}\, d\, to\, CN_s \\ & + \beta_{12} \operatorname{pr}(export)_{sj,lag} + \beta_{13}\, d\, to\, world_s + \beta_{14} imports\, from\, CN_s \\ & + \beta_{15} imports\, from\, world_s + FE + \mu(25) \end{aligned}$$

To test the impact of product heterogeneity on the number of firms in the industry (Proposition 1), the model takes the number of firms that an industry exports to a country as the dependent variable. The $N_s$ is not used in logarithm form because it is count data, and there are many zero observations (O'Hara and Kotze 2010). Since the number of firms is a discrete variable, the model uses Poisson Regression. The number of firms as dependent variables is shown as $N_{sk,now}$, the variable is not lagged form, but the term $N_{sk}$ is lagged for one year and is controlled in both this model and empirical model above.

In terms of explanatory variables, product homogeneity in the industry is the core explanatory variable. According to theory, the decline in product heterogeneity makes products similar in the industry, industry competition is more intense and production is more dispersed. Therefore, it is difficult for firm to pay the entry cost to enter the market, the number of firms entering foreign markets has fallen. Theoretically, the sign of coefficients of product heterogeneity is positive.

In terms of control variables, the control variables in this chapter are similar to the previous regressions, except that the interaction terms of the previous model have been removed. Clustered error at firm level is controlled.

Table 4 shows that industrial product heterogeneity has a significant positive impact on the number of enterprises in the industry. In the model, the coefficients of all variables are significant at the 1% significance level.

In terms of core explanatory variables, the results from the first to third columns gradually increase the control variables, but product heterogeneity always has a positive impact on the number of firms in a country. The decline in product heterogeneity means consumers cannot distinguish one product from another, and the market power of firms is smaller. This means that market competition is fierce, so that firms are small in size and cannot pay the entry cost, and therefore the number of firms which export to the nation is small (mechanism 1).

Table 4 The impact of product heterogeneity on number of firms in a market

|  | $N_{sk,now}$**** | $N_{sk,now}$ | $N_{sk,now}$ |
|---|---|---|---|
| $\gamma_k$ | 0.152*** | 0.160*** | 0.219*** |
|  | (0.00327) | (0.00515) | (0.0173) |
| $scale_j$ |  |  | -0.00137*** |
|  |  |  | (0.000458) |
| $\Sigma_{i,industry} d'_{isk}$ |  |  | -0.0345*** |
|  |  |  | (0.000145) |
| $markets_j$ |  |  | 0.00856*** |
|  |  |  | (0.000476) |
| $N_{sk}$ |  | 0.00802*** | 0.00767*** |
|  |  | (0.0000291) | (0.0000247) |
| $\Sigma_{i,firm} d'_{isj}$ |  | -0.00173*** | -0.00489*** |
|  |  | (0.0000485) | (0.000283) |
| $\Sigma_{i,industry} e^{-N_i/L_i+1} d'_{isk}$ |  |  | 0.0172*** |
|  |  |  | (0.000101) |
| $GDP_s$ |  | 0.0436*** | 0.0358*** |
|  |  | (0.000662) | (0.000542) |
| $d\ to\ CN_s$ |  | -0.122*** | 0.0820*** |
|  |  | (0.00221) | (0.00168) |
| $d\ to\ world_s$ |  | -0.000268*** | 0.0162*** |
|  |  | (0.0000158) | (0.0000701) |
| $imports\ from\ world_s$ |  | -0.0705*** | -0.0423*** |
|  |  | (0.000443) | (0.000533) |
| $imports\ from\ CN_s$ |  | 0.0586*** | 0.0615*** |
|  |  | (0.00125) | (0.00139) |
| N | 7073320 | 7005002 | 7005002 |
| Industry FE | Yes | Yes | Yes |
| Year FE | Yes | Yes | Yes |

\* $p < 0.1$, \*\* $p < 0.05$, \*\*\* $p < 0.01$. .\*\*\*\* N_(sk,now), the variable is not lagged form, but the term N_sk is lagged for one year. The clustered standard error is controlled at firm level.

Empirically, if industrial product heterogeneity increase, the number of firms in the industry at the market will increase, and this proves the Proposition 1. The above mechanism strongly explains why the increase in the number of firms leads to the increase in the probability of firms (Proposition 3), and it also shows the reason why firms prefer the markets where are less peers exist in the market (Proposition 4).

3.7 Robust test

The robustness test focuses on the impact of changes in the sample and model settings on the previous conclusions. The changes in the model settings or the samples have limited impact on the main conclusions.

First, the robustness test removed the samples with the largest and smallest 5% of the variable values of the number of firms ($N_{sk}$), main conclusions do not change.

In addition, the model adds national and regional fixed effects to control the factors of geography and international relation. Although the coefficients changed slightly, the main conclusions do not change (Table 5).

Firstly, the model also removes some control variables and adds country fixed effects. Country fixed effects control the stable variables given country, such as a country's economic scale, distance from China, and distance from other countries. Country fixed effects can substitute some variables in baseline model, and Probit regression cannot be estimated if some variables are put into the model due to multicollinearity. Therefore, in the third and fourth columns, the model remove the country-related variables, and use the country fixed effect to control these factors.

However, the country fixed effects control some new factors than the baseline model:

colonial relations, common language. These factors can affect the network relations between countries, and these relations reflect the influence of historical and international relations factors. If the country fixed effects make coefficient changes, which means that the historical and cultural ties between countries can impact exports.

The country fixed effect does not change main conclusions, except that the country fixed effect makes the coefficients smaller than those in the baseline regression. The historical and cultural ties between countries are related to the expansion of the network of industries and firms, and the ties between countries promote trade between firms. This is consistent with previous research of impact of national ties on national trade (Rauch 1999).

In addition, the model adds regional fixed effect, and the conclusions do not change. The regional fixed effect is added because the exports behavior of firms can be impacted by other firms in the same region (Zhang, Wang, and Li 2016). The robustness test adds the regional fixed effects, controlling the impact of communication of firms in the same region. The regional variable in the model is whether the firm belongs to a city. The regional fixed effect does not change the main conclusions (Table 5), except that the impact of product heterogeneity on the number of firms in the market becomes smaller. This shows that geographic agglomeration in the region has little impact on results in this model.

Table 5 Robust test of country fixed effect or region fixed effect

|  | $\Pr(export)_s$ | $N_{sk,now}$**** | $\Pr(export)_s$ | $N_{sk,now}$ |
|---|---|---|---|---|
| $N_{sk}$ | 0.00287*** | 0.00359*** | 0.00763*** | 0.00821*** |
|  | (0.000110) | (0.0000145) | (0.000120) | (0.0000246) |
| $\Sigma_{i,industry}d'_{isk}$ | -0.00318*** | -0.00663*** | -0.00419*** | -0.0171*** |
|  | (0.000214) | (0.0000585) | (0.000120) | (0.0000981) |
| $\Sigma_{i,industry}e^{-N_i/L_i+1}d'_{isk}$ | 0.00104*** | 0.00422*** | 0.00395*** | 0.0189*** |
|  | (0.000130) | (0.0000369) | (0.000132) | (0.0000937) |
| $\gamma N_s$ | -0.00143*** |  | -0.00198*** |  |
|  | (0.000121) |  | (0.000147) |  |
| $\gamma\Sigma_{i,industry}e^{-N_i/L_i+1}d'_{is}$ | 0.000455*** |  | 0.00202*** |  |
|  | (0.000148) |  | (0.000161) |  |
| $markets_i$ | 0.0463*** | 0.00299*** | 0.0307*** | 0.000994** |
|  | (0.000979) | (0.000167) | (0.000856) | (0.000418) |
| $\Sigma_{i,firm}d'_{isi}$ | -0.0185*** | -0.00171*** | -0.0132*** | -0.000543** |
|  | (0.000420) | (0.0000981) | (0.000373) | (0.000247) |
| $scale_i$ | 0.158*** | 0.000494*** | 0.160*** | -0.00206*** |
|  | (0.00291) | (0.000136) | (0.00269) | (0.000447) |
| $\gamma_k$ | 0.104** | 0.279*** | 0.0766 | 1.690*** |
| | (0.0518) | (0.00594) | (0.0522) | (0.0177) |
| $\text{pr}(export)_{s,laa}$ | 1.767*** | | 2.007*** | |
| | (0.00658) | | (0.00651) | |
| $imports\ from\ world_s$ | 0.103*** | 0.157*** | 0.0109*** | -0.0534*** |
| | (0.00640) | (0.000918) | (0.00236) | (0.000644) |
| $imports\ from\ CN_s$ | 0.126*** | 0.0416*** | 0.0653*** | 0.0426*** |
| | (0.00786) | (0.00200) | (0.00301) | (0.00120) |
| N | 6961427 | 7005002 | 7003066 | 7005002 |
| Region FE | No | No | Yes | Yes |
| Industry FE | Yes | Yes | Yes | Yes |
| Year FE | Yes | Yes | Yes | Yes |
| Country FE | Yes | Yes | No | No |

\* $p < 0.1$, \*\* $p < 0.05$, \*\*\* $p < 0.01$. .\*\*\*\* N_(sk,now), the variable is not lagged form, but the term N_sk is lagged for one year. The clustered standard error is controlled at firm level.

## 4. Conclusion

From the above analysis, industry export network can enter firstly in one country and enter its neighbors' markets. Firms can use information of other firms in the industry to enter the market, and search for nearby markets of existing markets of other firms. When the number of firms entering a certain market in a certain industry increases, the firms obtain more information about the market, which promotes other firms to enter the market. The information can also make firms to search for new markets around the market, but the market competition makes firms prefer new markets with fewer peers. In terms of mechanism, the number of firms in a market is positively affected by product heterogeneity. Heterogeneity products increase the market power of firms and the firms have the ability to pay for entry costs. This reflects the impact of distance between different markets on export of firms, which explains the economic significance of geographic distance in international trade. In addition, the reduction of product heterogeneity itself can increase reference value of peer information, which makes firms more willing to follow the market and their markets is closer with each other. In addition, the interactive influence of trade network among industry, country, and firm factors needs further studies.

# Reference

Bailey, Michael, Abhinav Gupta, Sebastian Hillenbrand, Theresa Kuchler, Robert J. Richmond, and Johannes Stroebel. 2020. International trade and social connectedness. In *Working Paper*.

Brandt, Loren, Johannes Van Biesebroeck, Luhang Wang, and Yifan Zhang. 2017. "WTO Accession and Performance of Chinese Manufacturing Firms." *American Economic Review* no. 107 (9):2784-2820. doi: 10.1257/aer.20121266.

Brandt, Loren, Johannes Van Biesebroeck, and Yifan Zhang. 2012. "Creative accounting or creative destruction? Firm-level productivity growth in Chinese manufacturing." *Journal of Development Economics* no. 97 (2):339-351. doi: https://doi.org/10.1016/j.jdeveco.2011.02.002.

Chaney, Thomas. 2014. "The Network Structure of International Trade." *American Economic Review* no. 104 (11):3600-3634.

De Lucio, Juan, Raúl Mínguez, Asier Minondo, and Francisco Requena. 2016. "Networks and the Dynamics of Firms' Export Portfolio: Evidence for Mexico." *The World Economy* no. 39 (5):708-736. doi: 10.1111/twec.12286.

Erbahar, Aksel. 2019. "Market knowledge: Evidence from importers." *The World Economy* no. 42 (4):1110-1151. doi: 10.1111/twec.12750.

Fan, Haichao, Edwin L. C. Lai, and Yao Amber Li. 2015. "Credit constraints, quality, and export prices: Theory and evidence from China." *Journal of Comparative Economics* no. 43 (2):390-416. doi: https://doi.org/10.1016/j.jce.2015.02.007.

Gao, Chao, Jiuli Huang, and Kunwang Li. 2019. "Dialect, history of immigrants and regional trade (In Chinese. With English summary.)." *Guanli Shijie /Mangement World Journal* no. 35 (02):43-57.

Hummels, David, and Peter J. Klenow. 2005. "The Variety and Quality of a Nation's Exports." *American Economic Review* no. 95 (3):704-723. doi: 10.1257/0002828054201396.

Melitz, Marc J. 2003. "The Impact of Trade on Intra-Industry Reallocations and Aggregate Industry Productivity." *Econometrica* no. 71 (6):1695-1725. doi: 10.1111/1468-0262.00467.

Melitz, Marc J., and Gianmarco I. P. Ottaviano. 2008. "Market Size, Trade, and Productivity." *The Review of Economic Studies* no. 75 (1):295-316. doi: 10.1111/j.1467-937X.2007.00463.x.

O'Hara, Robert B., and D. Johan Kotze. 2010. "Do not log‐transform count data." *Methods in Ecology & Evolution* no. 1.

Rauch, James. 2001. "Business and Social Networks in International Trade." *Journal of Economic Literature* no. 39:1177-1203. doi: 10.1257/jel.39.4.1177.

Rauch, James E. 1999. "Networks Versus Markets in International Trade." *Journal of International Economics* no. 48 (1):7-35.

Ruhl, Kim J., and Jonathan L. Willis. 2017. "NEW EXPORTER DYNAMICS." *International Economic Review* no. 58 (3):703-726. doi: 10.1111/iere.12232.

Wu, Qunfeng, and Rudai Yang. 2019. "Trade and Networks: a method of gravity model (In Chinese. With English summary.)." *Jingji Yanjiu/ Economic Research Journal* no. 54 (02):84-101.

Yang, Rudai, and Yan Li. 2016. "Immigrant networks and dynamics of trade(In Chinese. With English summary.)." *Jingji Yanjiu/ Economic Research Journal* no. 51 (03):163-175.

Zhang, Feng. 2014. *The impact of fix cost on extensive margin of exports (In Chinese. With English summary.)*. 博士, Shandong University.

Zhang, Guoreng, Yongjin Wang, and Kunwang Li. 2016. "Exports of firm and Industrial Cluster: the analysis of social and communicational factors(In Chinese)." *Shijie Jingji/ World Economy*

*Journal* no. 39 (02):48-74.

## Mathematical appendix:

The proof of equation (1.9)

$$\lim_{d_{os}\to\infty} \frac{\partial N_s}{\partial \gamma_j} = \frac{\frac{1}{\gamma_j^2}\sigma ie^{-N_s+1} + \frac{(1-\sigma)}{\gamma_j^2}}{-\sigma ie^{-N_s+1}\left(\eta N_s - \eta + \frac{1}{\gamma_j}\right) + (1-\sigma)\gamma_j\eta} \approx \frac{1}{\gamma_j^3\eta} > 0 \quad (1.9)$$

Proof: from (1.23) (1.24)(1.25)

$$I_3 = \sigma(ie^{-N_s+1} d_s) + (1-\sigma)\left(d_{0s} + d_s' - d_s^p\right) \tag{1.5}$$

$$c_{Ds} = I_3 + I_4 \tag{1.6}$$

$$I_4 = \frac{s_s}{\eta N_s + 1/\gamma_j}\left(\frac{\alpha}{\gamma_j} + \eta N_s \ \bar{p}\right) \tag{1.7}$$

$$c_{Ds} = \sigma(ie^{-N_s+1} d_s) + (1-\sigma)\left(d_{0s} + d_s' - d_s^p\right) + \frac{s_s}{\eta N_s + 1/\gamma_j}\left(\frac{\alpha}{\gamma_j} + \eta N_s \ \bar{p}\right) \tag{A.1}$$

$$0 = \left(\frac{\alpha s_s}{\gamma_j} + \eta s_s N_s \ \bar{p}\right) - \left(\eta N_s + \frac{1}{\gamma_j}\right)c_{Ds} + \sigma(ie^{-N_s+1} d_s)(\eta N_s + \frac{1}{\gamma_j})$$

$$+ (1-\sigma)\left(d_{0s} + d_s' - d_s^p\right)(\eta N_s + \frac{1}{\gamma_j}) \tag{A.2}$$

$$0 = \left(-\frac{\alpha s_s}{\gamma_j^2} d\gamma_j + \eta s_s \bar{p} dN_s\right) - \left(\eta dN_s - \frac{1}{\gamma_j^2} d\gamma_j\right) c_{Ds} - \sigma ie^{-N_s+1} d_s\left(\eta N_s + \frac{1}{\gamma_j}\right) dN_s$$

$$+ \sigma ie^{-N_s+1} d_s \left(\eta dN_s - \frac{1}{\gamma_j^2} d\gamma_j\right)$$

$$+ (1-\sigma)\left(d_{0s} + d_s' - d_s^p\right)\left(\eta dN_s - \frac{1}{\gamma_j^2} d\gamma_j\right) \tag{A.3}$$

$$0 = \left(\frac{c_{DS}}{\gamma_j^2} - \frac{\alpha s_s}{\gamma_j^2} - \frac{1}{\gamma_j^2}\sigma ie^{-N_s+1} d_s - \frac{(1-\sigma)}{\gamma_j^2}\left(d_{0s} + d_s' - d_s^p\right)\right) d\gamma_j$$

$$+ \left(\eta s_s\bar{p} - \eta c_{Ds} - \sigma ie^{-N_s+1} d_s \left(\eta N_s + \frac{1}{\gamma_j}\right) + \sigma ie^{-N_s+1} d_s \eta\right.$$

$$\left. + (1-\sigma)\eta\left(d_{0s} + d_s' - d_s^p\right)\right) dN_s \tag{A.4}$$

$$\frac{\partial N_s}{\partial \gamma_j}$$

$$= \frac{-\frac{c_{Ds}}{\gamma_j^2} + \frac{\alpha s_s}{\gamma_j^2} + \frac{1}{\gamma_j^2}\sigma i e^{-N_s+1} d_s + \frac{(1-\sigma)}{\gamma_j^2}(d_{0s} + d_s' - d_s^\rho)}{\eta s_s \bar{p} - \eta c_{Ds} - \sigma i e^{-N_s+1} d_s \left(\eta N_s + \frac{1}{\gamma_j}\right) + \sigma i e^{-N_s+1} d_s \eta + (1-\sigma)(d_{0s} + d_s' - d_s^\rho)\eta} \quad (A.5)$$

$$\lim_{d_{0s}\to\infty} \frac{\partial N_s}{\partial \gamma_j} = \frac{\frac{1}{\gamma_j^2}\sigma i e^{-N_s+1} + \frac{(1-\sigma)}{\gamma_j^2}}{-\sigma i e^{-N_s+1}\left(\eta N_s - \eta + \frac{1}{\gamma_j}\right) + (1-\sigma)\gamma_j \eta} \approx \frac{1}{\gamma_j^3 \eta} > 0 \quad (1.9)$$

The proof of 1.21, 1.22

$$\frac{\partial^2 c_{Ds}}{\partial \gamma_j \partial N_s} = \frac{-s_s}{(\eta N_s \gamma_j + 1)^4}\left((\alpha\eta - \eta\bar{p})\right)\left(-\eta^2 N_s^2 \gamma_j^2 + 1\right) < 0 \quad (1.21)$$

$$\frac{\partial^2 c_{Ds}}{\partial \gamma_j \partial d_s^\rho} = \frac{\partial^2 c_{Ds}}{\partial \gamma_j \partial N_s}\frac{dN_s}{dd_s^\rho} = \frac{1}{\Sigma_i e^{-N_i/L_i+1}/L_i d_{is}'}\frac{s_s}{(\eta N_s + \gamma_j)^4}\left((\alpha\eta - \eta\bar{p})\right)(-\eta^2 N_s^2 \gamma_j^2 + 1) > 0 \quad (1.22)$$

Proof:
From (1.5)(1.6)(1.7)

$$I_3 = \sigma(i e^{-N_s+1} d_s) + (1-\sigma)\left(d_{0s} + d_s' - d_s^\rho\right) \quad (1.5)$$

$$c_{Ds} = I_3 + I_4 \quad (1.6)$$

$$I_4 = \frac{s_s}{\eta N_s + 1/\gamma_j}\left(\frac{\alpha}{\gamma_j} + \eta N_s \bar{p}\right) \quad (1.7)$$

$$I_4 = \frac{s_s}{\eta N_s \gamma_j + 1}(\alpha + \eta \gamma_j N_s \bar{p}) \quad (A.6)$$

$$\frac{\partial c_{Ds}}{\partial \gamma_j} = s_s \frac{\left(\eta N_s \bar{p}(\eta N_s \gamma_j + 1) - \eta N_s(\alpha + \eta \gamma_j N_s \bar{p})\right)}{(\eta N_s \gamma_j + 1)^2} \quad (A.7)$$

$$\frac{\partial c_{Ds}}{\partial \gamma_j} = \frac{s_s(\eta N_s \bar{p} - \eta N_s \alpha)}{(\eta N_s \gamma_j + 1)^2} \quad (A.8)$$

$$\frac{\partial^2 c_{Ds}}{\partial \gamma_j \partial N_s} = \frac{s_s}{(\eta N_s \gamma_j + 1)^4}\left((-\alpha\eta + \eta\bar{p})(\eta N_s \gamma_j + 1)^2 - 2\eta\gamma_j(\eta N_s \gamma_j + 1)(-\alpha N_s + \eta N_s \bar{p})\right) \quad (A.9)$$

From (1.2), $\alpha > \bar{p}$

$$\frac{\partial^2 c_{Ds}}{\partial \gamma_j \partial N_s} = \frac{-s_s}{(\eta N_s + \gamma_j)^4}((\alpha\eta - \eta\overline{p}))(-\eta^2 N_s^2 \gamma_j^2 + 1) < 0 \qquad (1.21)$$

From (1.18)

$\Sigma_i e^{-N_i/L_i+1} d'_{is} = d_s^\rho$ （1.16） given $d'_{is}$, L

$$\frac{\partial^2 c_{Ds}}{\partial \gamma_j \partial d_s^\rho} = \frac{\partial^2 c_{Ds}}{\partial \gamma_j \partial N_s}\frac{dN_s}{dd_s^\rho} = \frac{1}{\Sigma_i e^{-N_i/L_i+1}/L_i d'_{is}}\frac{s_s}{(\eta N_s + \gamma_j)^4}((\alpha\eta - \eta\overline{p}))(-\eta^2 N_s^2 \gamma_j^2 + 1) > 0 \quad (1.22)$$